\documentclass[12pt]{article}
\usepackage[dvipsnames]{xcolor}
\usepackage{pgfplots}
\pgfplotsset{width=7cm,compat=1.17}
\usetikzlibrary{patterns}

\newcommand{\be}{\begin{equation}}
\newcommand{\ee}{\end{equation}}
\newcommand{\bey}{\begin{eqnarray}}
\newcommand{\pslash}{\not{\hbox{\kern-2.3pt $p$}}}
\newcommand{\pdslash}{\not{\hbox{\kern-2pt $\partial$}}}

\newcommand{\beq}{{\begin{equation}}}
\newcommand{\eeq}{{\end{equation}}}
\newcommand{\bea}{{\begin{array}}}
\newcommand{\eea}{{\end{array}}}
\newcommand{\ri}{{\rm i}}

\newcommand{\eey}{\end{eqnarray}}

\begin{document}

\begin{titlepage}
\vskip 2cm
\begin{center}
{\Large\bf Rest energy effects for non-relativistic Fermi fields in external potentials}
\vskip 3cm
{\bf
Fuad M. Saradzhev}
\footnote{{\tt fuads@athabascau.ca} }
\vskip 5pt
{\sl Centre for Science, Athabasca University,
 Athabasca, Alberta, Canada \\}
\vskip 2pt

\end{center}
\vskip .5cm
\rm

\begin{abstract}

The generalized L\'{e}vy-Leblond equation (GLL) is used to study the bound state problem for a non-relativistic Fermi field in external potentials. A condition when the rest energy term can be removed from the GLL equation is determined. For external potentials which do not obey the condition, the bound state spectrum is affected non-trivially by the rest energy of the Fermi field. This is demonstrated for a spherical finite-depth well potential. The existence, number, and energies of bound states depend on the value of the rest energy.

\end{abstract}

\vspace{1 cm}


\end{titlepage}

\setcounter{footnote}{0} \setcounter{page}{1} \setcounter{section}{0} %
\setcounter{subsection}{0} \setcounter{subsubsection}{0}

\section{Introduction}

Non-relativistic particles (fields), defined through the theory of representations of the Galilei group, are known
to acquire spin and magnetic moment \cite{levy}-\cite{lek}, the properties which were originally believed to be purely relativistic. The dynamics of the spin $1/2$ particles is governed by the L\'{e}vy-Leblond (LL) equation, the non-relativistic version of the Dirac equation. In \cite{levy}, it was derived in the Dirac way by linearization of the Schr\"{o}dinger equation.

Another way of deriving the LL equation is by making use of a $(4+1)$-dimensional covariant formulation of Galilean covariance \cite{ap1999}-\cite{sun2} based on an extended space-time approach \cite{takahashi}-\cite{omote}. The idea of such an approach in the non-relativistic framework appeared first in \cite{eisen} and then rediscovered in \cite{duval1985}. It was used to study the symmetries of the LL equation in the non-relativistic Chern-Simons electrodynamics \cite{duval1996}. A formulation of Galilean covariance in one higher dimension allows us to incorporate into the theory the notion of rest (internal) energy: one of the Casimir invariants of the Galilei group in $(4+1)$-dimensions includes the rest energy operator. Reduction to $(3+1)$ dimensions by eliminating the additional, fifth coordinate results in the generalized LL equation with a rest energy contribution \cite{sar2011}. A generalized version of the LL equation was also discussed in \cite{hueg2013}.

It was shown in \cite{sour} that the rest energy is an additive constant related to the particular structure of the Galilei group and that the relativistic energy does not contain such constant. Any additive energy constant including the rest energy can be considered as a reference point for the energy of non-relativistic systems, and the only effect of taking this constant into account is a shift of all energy levels by its value. This is equivalent to a phase change in the wave function. In this paper, we study the GLL equation in external potentials, and we aim to show that for some external potentials the assumption of rest energy additivity is violated. The rest energy term cannot be removed by a shift of energy levels or by redefining the Fermi field components. 

The physical consequences of rest energy in non-relativistic physics have not been studied before. Our motivation is to reveal the dynamical effects of rest energy, in particular, its non-trivial impact on the bound state spectrum of non-relativistic Fermi particles.

Our paper is organized as follows. In Sect.2, we consider a free non-relativistic Fermi field. We define a transformation for the Fermi field which removes the rest energy term from the GLL equation. The transformation is a combination of phase change
and mixing the field components. This demonstrates the equivalence of the GLL and LL equations for free non-relativistic Fermi fields.

We establish a two-step transition from the Dirac equation to the GLL equation. Taking the non-relativistic limit of the Dirac equation yields the LL equation \cite{levy}, and an additional transformation is required to get its generalized version.

In Sect.3, we consider a non-relativistic Fermi field in external potentials. We find that for some types of potentials the rest energy cannot be removed from the GLL equation. We determine the condition under which the rest energy can be removed. The component form of the GLL equation with an external potential is given. 

As an example of external potential for which the condition does not hold, we study in detail the case of a potential of spherical well type of finite depth. We derive a Schr\"{o}dinger type equation for the independent component of the Fermi field with an effective potential exhibiting the rest energy dependence. We determine the spectrum of the bound S-states, and we find the minimum value of the rest energy below which no bound states exist.

We conclude with Discussion in Sect.4.

\section{GLL equation for a free non-relativistic Fermi field}

The Schr\"{o}dinger equation for a free non-relativistic Fermi field ${\psi}({\bf r},t)$ with inertial mass $m$ is
\begin{equation}
{\rm i} {\hbar} \frac{{\partial}{\psi}({\bf r},t)}{{\partial}t} =
\left( - {\hbar}^2 \frac{\Delta}{2m} + E_0 \right) {\psi}({\bf r},t),
\label{schr1}
\end{equation}
where ${\bf r}=(x^1=x, x^2=y, x^3=z)$, ${\Delta} \equiv {\partial}_a^2$, ${\partial}_a \equiv \frac{\partial}{{\partial}x^a}$, $a=1,2,3$, and $E_0$ is an arbitrary constant energy.

By re-defining the wave function ${\psi}({\bf r},t)$ as
\begin{equation}
{\psi}({\bf r},t) \to \tilde{\psi}({\bf r},t) \equiv e^{\frac{\rm i}{\hbar} E_0 t} {\psi}({\bf r},t),
\label{redef}
\end{equation}
we can remove the constant energy term from the Schr\"{o}dinger equation:
\begin{equation}
{\rm i} {\hbar} \frac{{\partial}{\tilde{\psi}}({\bf r},t)}{{\partial}t} =
 - {\hbar}^2 \frac{\Delta}{2m} \tilde{\psi}({\bf r},t).
\label{eqtri}
\end{equation}

The GLL equation for a $4$-component non-relativistic Fermi field
${\psi}({\bf r},t)$ reads \cite{sar2011}
\begin{equation}
{\rm i} {\hbar}{\gamma}^4 \frac{{\partial}{\psi}({\bf r},t)}{{\partial}t} =
\left(-{\rm i}{\hbar} c
\gamma^{a}\partial_{a} + c kI_{+} \right){\psi}({\bf r},t)
\label{becompact}
\end{equation}
or
\begin{equation}
\left({\rm i}{\hbar}
\gamma^{\bar{\mu}}\partial_{\bar{\mu}}-kI_{+} \right){\psi}({\bf r},t)=0,
\label{compact}
\end{equation}
where $\bar{\mu}$ runs from $1$ to $4$, ${\partial}_4 \equiv \frac{1}{c} \frac{\partial}{{\partial}t}$ and
\begin{equation}
I_{+} \equiv I - \frac{m{c}}{k} {\gamma}^5,
\label{specmat}
\end{equation}
$I$ being the identity matrix, and $k$ being the momentum corresponding to the rest energy $E_{\rm rest} = \frac{k^2}{2m}$.
For $k=0$, Eq(\ref{compact}) reduces to the LL equation.

The ${\gamma}$-matrices
\begin{eqnarray}
\gamma^a=
\left(\begin{array}{cc}
0 & \ri\sigma^a\\
\ri\sigma^a & 0
\end{array}\right),\;\;\;
\gamma^{4}=\frac{1}{\sqrt{2}} \left(
\begin{array}{cc}
1 & 1 \\
-1 & -1
\end{array}\right), 
\nonumber
\end{eqnarray}
\begin{eqnarray}
\gamma^5=\frac{1}{\sqrt{2}} \left(
\begin{array}{cc}
1 & -1\\
1 & -1
\end{array}\right) 
\end{eqnarray}
satisfy the algebra
\begin{equation}
\gamma^{\mu} \gamma^{\nu} + \gamma^{\nu} \gamma^{\mu} = 2g^{\mu \nu},
\label{gammaalgebra}
\end{equation}
with ${\mu},{\nu}=1,...,5$ and
\begin{equation}
g^{\mu\nu}=\left( \begin{array}{ccc} -{\mathbf 1}_{3\times 3}
& 0 & 0 \\ 0 & 0 & 1 \\ 0 & 1 & 0\end{array} \right).
\end{equation}

We could consider the GLL equation with a constant energy term as well. In Eq.(\ref{becompact}), this term would be ${\gamma}^4 E_0$. As in the case of the Schr\"{o}dinger equation, it could be removed from the GLL equation by using the re-definition given by Eq.(\ref{redef}). 

The rest energy term can also be removed from the GLL equation. To show this, let us introduce the new field
\begin{equation}
{\psi}({\bf r},t) \to {\eta}({\bf r},t)
\equiv
\left( I - \frac{{\mu}}{\sqrt{2}} {\gamma}^4 \right) e^{\frac{\rm i}{\hbar} E_{\rm rest} t} {\psi}({\bf r},t),
\end{equation}
where ${\mu}$ is a parameter to be determined. The GLL equation given by Eq.(\ref{compact}) becomes
\begin{equation}
\Big[{\rm i}{\hbar}
\gamma^{4}\partial_{4} + {\rm i}{\hbar} \left( I - \frac{{\mu}}{\sqrt{2}} {\gamma}^4 \right) {\gamma}^a {\partial}_a 
-kI_{+}\left( I + \frac{{\mu}}{\sqrt{2}}{\gamma}^4 \right) + \frac{1}{c} E_{\rm rest} {\gamma}^4 \Big]{\eta}({\bf r},t)=0.
\end{equation}
Multiplying both sides of this equation from the left by $\left( I + \frac{{\mu}}{\sqrt{2}} {\gamma}^4 \right)$, we bring it to the form
\begin{equation}
\Big[{\rm i}{\hbar}
\gamma^{\bar{\mu}}\partial_{\bar{\mu}} + m{c} {\gamma}^5 + \left( {\mu} m{c} \sqrt{2} - k \right)I
+ \frac{1}{2m{c}} {\left( {\mu} m{c} \sqrt{2} - k \right)}^2 {\gamma}^4 \Big]{\eta}({\bf r},t)=0.
\label{newform}
\end{equation}
If we take
\begin{equation}
{\mu} = \frac{k}{m{c}\sqrt{2}} = \sqrt{\frac{m_0}{m}},
\label{newparameter}
\end{equation}
where $m_0$ is a rest mass associated with the rest energy as $m_0 \equiv E_{\rm rest}/c^2$, then the last two terms in Eq.(\ref{newform}) vanish, so the field
\begin{equation}
{\eta}({\bf r},t)
= \left( I - \sqrt{\frac{m_0}{2m}} {\gamma}^4 \right) e^{\frac{\rm i}{\hbar} E_{\rm rest} t} {\psi}({\bf r},t)
\label{newfield}
\end{equation}
obeys the original LL equation without the rest energy term:
\begin{equation}
\left({\rm i}{\hbar}
\gamma^{\bar{\mu}}\partial_{\bar{\mu}} + m{c} {\gamma}^5 \right) {\eta}({\bf r},t)=0.
\label{original}
\end{equation}
In addition to the exponential time-dependent factor which we had before in Eq.(\ref{redef}), we get here the factor $( 1 - \sqrt{\frac{m_0}{2m}}{\gamma}^4)$ that depends on the ratio of the rest and inertial masses and mixes the components of the original field  ${\psi}({\bf r},t)$.

Representing the field ${\eta}({\bf r},t)$ as
\begin{equation}
{\eta}({\bf r},t)=
\left(
\begin{array}{c}
{\eta}_{1}({\bf r},t) \\ {\eta}_{2}({\bf r},t)
\end{array}
\right),
\end{equation}
where ${\eta}_{1}({\bf r},t)$, ${\eta}_{2}({\bf r},t)$ are $2$-component fields, and introducing their linear combinations
\begin{eqnarray}
{\eta}_{+}({\bf r},t) & \equiv & {\eta}_{1}({\bf r},t) + {\eta}_{2}({\bf r},t), \nonumber \\
{\eta}_{-}({\bf r},t) & \equiv & {\eta}_{1}({\bf r},t) - {\eta}_{2}({\bf r},t), 
\end{eqnarray}
we can rewrite the GLL equation as a system of two equations:
\begin{eqnarray}
{\hbar} \sqrt{2}  {\partial}_4 {\eta}_{+} +  (\mbox{\boldmath ${\sigma}$} {\bf p})  {\eta}_{-} & = & 0, \nonumber \\
{\rm i}  (\mbox{\boldmath ${\sigma}$} {\bf p}) {\eta}_{+} - m{c} \sqrt{2} {\eta}_{-} & = & 0.
\label{system1}
\end{eqnarray}
where ${\bf p} = - {\rm i} {\hbar} {\bf {\nabla}} = - {\rm i} {\hbar} ({\partial}_1,{\partial}_2,{\partial}_3)$.

Only one of the components is dynamically independent. Its time evolution determines the time evolution of another component as well. Eliminating, for instance, ${\eta}_{-}({\bf r},t)$ in favor of ${\eta}_{+}({\bf r},t)$, this brings us back to the Schr\"{o}dinger equation without the rest energy term:
\begin{equation}
{\rm i} {\hbar} \frac{{\partial}{{\eta}_{+}}({\bf r},t)}{{\partial}t} =
 - {\hbar}^2 \frac{\Delta}{2m} {\eta}_{+}({\bf r},t).
\label{schr2}
\end{equation}
The GLL equation for a free non-relativistic Fermi field can be therefore reduced to the $2$ x $2$ matrix Schr\"{o}dinger equation for its independent component. In this sense, the GLL equation and the Schr\"{o}dinger equation are equivalent to each other. However, starting with the GLL equation allows us to study both the spin and rest energy effects. 

The GLL equation can be connected with the Dirac equation as well. The rest energy of a free relativistic Fermi field is incorporated nontrivially in its total
or relativistic energy. Its rest mass $m_0$ is at the same time a measure of its inertia, i.e. of
the tendency of the field to resist changes in velocity. Thus, $m_0$ is its inertial mass as well
\cite{jam}.

For non-relativistic Fermi fields, the rest energy contribution decouples from the kinetic energy
and becomes an additive factor. In this case, the rest
energy is independent from inertia, while the rest mass $m_0$ is not in general the same as the inertial
mass $m$ \cite{jack},\cite{pa} and can be interpreted as a parameter related to intrinsic structure of Fermi particles. 

In the limit $c \to \infty$, the rest energy becomes infinitely large in value and should be subtracted
from the total energy before the non-relativistic limit of the Dirac equation is taken. This is achieved
by re-defining the Dirac wave function ${\psi}_{D}({\bf r},t)$ in the way similar to Eq.(\ref{redef}):
\begin{equation}
{\psi}_{D}({\bf r},t) \to \tilde{\psi}_{D}({\bf r},t) \equiv e^{\frac{\rm i}{\hbar} m_0 c^2 t} {\psi}_{D}({\bf r},t).
\label{redef1}
\end{equation}
As a result, the non-relativistic limit of the Dirac equation is the LL equation without any rest energy
contribution \cite{levy}. It is sometimes stated that the rest energy is a purely relativistic effect and has no
non-relativistic limit \cite{tha}.

The transition from the Dirac equation to the GLL equation requires two steps. First, the
non-relativistic limit of the Dirac equation is taken. Then, within the non-relativistic formulation,
a final amount of rest energy $E_{\rm rest}$ is introduced by transforming the Fermi fields as
\begin{equation}
{\eta}({\bf r},t) \to {\psi}({\bf r},t) = 
\left( I + \sqrt{\frac{m_0}{2m}} {\gamma}^4 \right) e^{-\frac{\rm i}{\hbar} E_{\rm rest} t} {\eta}({\bf r},t).
\label{inverse}
\end{equation}
This transformation is the inverse of the transformation given by Eq.(\ref{newfield}).

\section{GLL equation with external potentials}

As in the case of the Dirac equation, external potentials can be included in the GLL equation to describe interactions of Fermi fields with external sources. A general form of the external potential can be represented by a $4 \times 4$ matrix-valued function $V({\bf r},t)$, so that the GLL equation becomes
\begin{equation}
\left({\rm i}{\hbar}
\gamma^{\bar{\mu}}\partial_{\bar{\mu}}-kI_{+} - V \right){\psi}({\bf r},t)=0.
\label{compactpot}
\end{equation}
Using again the field ${\eta}({\bf r},t)$ given by Eq.(\ref{newfield}), we convert Eq.(\ref{compactpot}) into
\begin{equation}
\left({\rm i}{\hbar}
\gamma^{\bar{\mu}}\partial_{\bar{\mu}} + m{c} {\gamma}^5 - \tilde{V} \right) {\eta}({\bf r},t)=0,
\label{originalpot}
\end{equation}
where
\begin{equation}
\tilde{V} \equiv V + \frac{1}{\sqrt{2}} {\mu} \left\{ {\gamma}^4, V \right\} +
\frac{1}{2} {\mu}^2 {\gamma}^4 \left\{ {\gamma}^4, V \right\},
\label{newpot}
\end{equation}
and $\left\{ A , B \right\} = AB + BA$ is the anticommutator of two operators $A$ and $B$.

If the external potential obeys the condition
\begin{equation}
\left\{ {\gamma}^4, V \right\} = 0,
\label{conditionpot}
\end{equation}
then $\tilde{V} = V$ and the rest energy term is removed from the GLL equation in the way similar to the case of free non-relativistic Fermi fields.

An example of such external potential is
\begin{equation}
V = \frac{1}{c} {\gamma}^4 {\varphi},
\label{pot1}
\end{equation}
where ${\varphi({\bf r},t)}$ is a single-valued function. It has the same matrix structure as the constant energy term ${\gamma}^4 E_0$. This results in the shift
\begin{equation}
{\rm i} {\hbar} {\gamma}^4  {\partial}_4 \to \frac{1}{c} {\gamma}^4 \left( {\rm i} {\hbar} \frac{\partial}{{\partial}t} - {\varphi} \right),
\end{equation}
indicating that  ${\varphi({\bf r},t)}$ in Eq.(\ref{pot1}) can be interpreted as potential energy (or potential function) of the system. For the hydrogen atom, ${\varphi}$ is the potential energy of the Coulomb interaction with no rest energy contribution \cite{ll2}, \cite{mita}.

The condition (\ref{conditionpot}) also holds for the potential produced by an electromagnetic field $A_{\bar{\mu}}$. It can be introduced by replacing ${\partial}_{\bar{\mu}}$ with $D_{\bar{\mu}} \equiv {\partial}_{\bar{\mu}} -
\frac{\rm i}{\hbar} \frac{e}{c} A_{\bar{\mu}}$, where $e$ is the charge of the Fermi field. This yields
\begin{equation}
V = - \frac{e}{c} {\gamma}^{\bar{\mu}} A_{\bar{\mu}}.
\label{em}
\end{equation}
The matrices ${\gamma}^{\bar{\mu}}$ anti-commute with ${\gamma}^4$, and the rest energy $E_{\rm rest}$ does not show itself in the solutions of Eq.(\ref{originalpot}). The GLL equation with the
potential given by Eq.(\ref{em}) is the linearized version of the Pauli equation \cite{gre},\cite{wilk}.

An example of potential that does not obey the condition (\ref{conditionpot}) is 
\begin{equation}
V = \frac{1}{c} {\varphi}.
\label{simplepot}
\end{equation}
The transformed potential
\begin{equation}
\tilde{V} = \frac{1}{c} \left( I + {\mu} \sqrt{2} {\gamma}^4 \right) {\varphi}
\end{equation}
exhibits the rest energy contribution in its second term. It means that for this type of potentials the rest energy will contribute non-trivially to the bound state spectrum of the GLL equation.

In the component form, Eq.(\ref{originalpot}) with the potential term given by Eq.(\ref{simplepot}) becomes
\begin{eqnarray}
{\rm i} {\hbar} \sqrt{2} \frac{{\partial} {\eta}_{+}}{{\partial}t}  +   \left[ {\rm i} c  (\mbox{\boldmath ${\sigma}$} {\bf p}) - {\varphi} \right] {\eta}_{-}  - 2{\mu} {\varphi} {\eta}_{+} & = & 0, \nonumber \\
\left[ {\rm i} c (\mbox{\boldmath ${\sigma}$} {\bf p}) + {\varphi} \right] {\eta}_{+} - mc^2 \sqrt{2} {\eta}_{-} & = & 0.
\label{systemext}
\end{eqnarray}
For time-independent and spherically symmetric external potentials, 
${\varphi} = {\varphi}(r)$, where $r = |{\bf r}|$.

Let us represent ${\eta}_{+}({\bf r},t)$ and  ${\eta}_{-}({\bf r},t)$ as
\begin{equation}
{\eta}_{\pm}({\bf r},t) = {\eta}_{\pm}({\bf r}) e^{-\frac{\rm i}{\hbar} Et}
\end{equation}
and introduce the dimensionless variables
\begin{equation}
\mbox{\boldmath $\xi$} \equiv \frac{m{c}}{\hbar} {\bf r}, \qquad {\cal E} \equiv \frac{E}{m{{c}}^2},
\qquad \bar{\varphi} \equiv \frac{\varphi}{m{c}^2}.
\end{equation}
Let us choose $\varphi$ in Eq.(\ref{simplepot}) in the form of the spherical well potential
\begin{equation}
\bar{\varphi}({\xi}) =
\left\{
\begin{array}{cc}
- \bar{\varphi}_0 & {\rm for} \hspace{5 mm} {\xi}<{\xi}_0,\\
0 & {\rm for} \hspace{5 mm} {\xi}>{\xi}_0,
\end{array}
\label{sphwello}
\right.
\end{equation}
where $\bar{\varphi}_0>0$. Then Eq.(\ref{systemext}) can be rewritten as the eigenvalue equation
\begin{equation}
{\cal H} {\eta}_{+}(\mbox{\boldmath $\xi$}) = {\cal E} {\eta}_{+}(\mbox{\boldmath $\xi$})
\label{eigeneq}
\end{equation}
with the effective Hamiltonian density
\begin{equation}
{\cal H} =  \frac{1}{2}  {\bf p_{\xi}^2} + V_{\rm eff},
\label{hamden}
\end{equation}
where  ${\bf p_{\xi}} = -{\rm i}{\bf {\nabla}_{\xi}} = - {\rm i} ({\hbar}/mc) {\bf {\nabla}}$. The shape of the effective potential 
\begin{equation}
V_{\rm eff}  =
\left\{
\begin{array}{cc}
- V_0  &  {\rm for} \qquad
{\xi}<{\xi}_0,\\
0 &  {\rm for} \qquad {\xi}>{\xi}_0,
\end{array}
\label{sphhamil}
\right.
\end{equation}
where 
\begin{equation}
V_0 = {\mu} \sqrt{2} \bar{\varphi}_0 - \frac{1}{2} {\bar{\varphi}_0}^2,
\end{equation}
is ${\mu}$-dependent. For
\begin{equation}
{\mu}>{\mu}_0 \equiv \frac{1}{2\sqrt{2}}  \bar{\varphi}_0,
\label{fcon}
\end{equation}
$V_{\rm eff}$ is a spherical well potential of depth $V_0$, while for ${\mu}<{\mu}_0$, including the case of the LL equation (${\mu}=0$), it becomes a potential step. For
\begin{equation}
{\mu} = {\mu}_0 + \frac{1}{\sqrt{2}},
\label{orig}
\end{equation}
the effective potential coincides with the original well potential given by Eq.(\ref{sphwello}).

In what follows, we assume that ${\mu}>{\mu}_0$, so that we can search for bound states. The spin terms do not appear in the effective Hamiltonian density. The set of operators commuting with ${\cal H}$ is ${\bf L}^2$,${\rm L}_{3}$, so the solutions of the Eq.(\ref{eigeneq}) have the form
\begin{equation}
{\eta}_{+,lM} = \frac{1}{\xi} R_{l} Y_{l}^{M},
\label{eigstates}
\end{equation}
where $\frac{1}{\xi} R_{l}$  and $Y_{l}^{M}$ are their radial and angular parts, respectively, $Y_{l}^{M}$ being spherical harmonics.

The quantum numbers $l, M$ are defined as
\begin{eqnarray}
{\bf L}^2  {\eta}_{+,lM} & = & l(l+1) {\eta}_{+,lM},\nonumber \\
{\rm L}_{3} {\eta}_{+,lM} & = & M {\eta}_{+,lM}.
\end{eqnarray}
Substituting the ansatz (\ref{eigstates}) into Eq.(\ref{eigeneq}) and using the expression for
${\bf p_{\xi}^2}$ in spherical polar coordinates
\begin{equation}
{\bf p_{\xi}^2} = - \frac{{\partial}^2}{{\partial}{\xi}^2} - \frac{2}{\xi} \frac{\partial}{{\partial}{\xi}} +
\frac{1}{{\xi}^2} {\bf L}_{\xi}^2,
\end{equation}
we bring the radial part of the eigenvalue equation to the form
\begin{equation}
\frac{d^2R_{l}}{d{\xi}^2} + \left( a^2 - \frac{l(l+1)}{{\xi}^2}
\right) R_{l} = 0 
\label{firsteq}
\end{equation}
for ${\xi}<{\xi}_0$, and
\begin{equation}
\frac{d^2R_{l}}{d{\xi}^2} + \left( - b^2 - \frac{l(l+1)}{{\xi}^2}
\right) R_{l} = 0 
\label{secondeq}
\end{equation}
for  ${\xi}>{\xi}_0$, where we have introduced the abbreviations
\begin{equation}
a^2 \equiv 2 \left( V_0 - |{\cal E}| \right),  \qquad  b^2 \equiv 2|{\cal E}|.
\label{abbre}
\end{equation}

We are interested in the bound state solutions for which $|{\cal E}|< V_0$.
The solutions are \cite{flugge},\cite{capri}
\begin{equation}
R_{l}({\xi}) =
\left\{
\begin{array}{cc}
N_{-} j_{l}(a{\xi}) &  {\rm for} \qquad
{\xi}<{\xi}_0,\\
N_{+} h_{l}^{(1)}({\rm i}b{\xi}) &  {\rm for} \qquad {\xi}>{\xi}_0,
\end{array}
\label{possolutions}
\right.
\end{equation}
where $N_{-}$,$N_{+}$ are normalization constants, while $j_{l}$ and $h_{l}^{(1)}$ are spherical Bessel and Hankel functions, respectively.
At ${\xi}={\xi}_0$, both $R_{l}({\xi})$ and its first derivative should be continuous. The normalization constants can be eliminated from the continuity relation of the logarithmic derivative. This yields the condition
\begin{equation}
{\rm i}b \frac{h_{l}^{(1){\prime}}({\rm i}b{\xi}_0)}
{h_{l}^{(1)}({\rm i}b{\xi}_0)} =
a \frac{j_{l}^{\prime}(a{\xi}_0)}{j_{l}(a{\xi}_0)},
\label{condition}
\end{equation}
where the primes denote differentiations to the respective arguments. This is the transcendental equation for the eigenvalue ${\cal E}$. The quantum number $M$ does not affect the eigenvalues, so each energy level is $(2l+1)$-fold degenerate.

For $l=0$, i.e. for S-states, the condition (\ref{condition}) reduces to
\begin{equation}
{\rho} {\rm cot}(\bar{\xi}_0 {\rho}) = - \sqrt{1 - {\rho}^2},
\label{simcondition}
\end{equation}
where
\begin{equation}
{\rho} \equiv \sqrt{1 - \frac{|{\cal E}|}{V_0}},
\qquad
\bar{\xi}_0 \equiv \sqrt{2V_0} {\xi}_0.
\end{equation}
The bound states spectrum is given by
\begin{equation}
|{\cal E}| = V_0 (1 - {\rho}^2),
\qquad
0 < {\rho} < 1.
\label{boundst}
\end{equation}
\begin{figure}[hbtp]
\vspace{-2cm}
\begin{picture}(200,400)
\put (10,320){\line(1,0){192}}
\put (10,280){\line(1,0){192}}
\put (10,210){\line(1,0){192}}
\put (202,210){\line(0,1){130}}
\put (154,210){\line(0,1){130}}
\put (106,210){\line(0,1){130}}
\put (58,210){\line(0,1){130}}
\put (10,210){\line(0,1){110}}
\put (58,340){\line(1,0){144}}

\put (82,330){\makebox(0,0){1}}
\put (130,330){\makebox(0,0){2}}
\put (178,330){\makebox(0,0){3}}

\put (82,300){\makebox(0,0){4}}
\put (130,300){\makebox(0,0){8}}
\put (178,300){\makebox(0,0){12}}

\put (34,300){\makebox(0,0){$V_0$}}
\put (34,250){\makebox(0,0){${\rho}$}}

\put (82,250){\makebox(0,0){0.765}}
\put (178,250){\makebox(0,0){0.479}}

\put (82,265){\makebox(0,0){0.388}}
\put (130,265){\makebox(0,0){0.285}}
\put (130,250){\makebox(0,0){0.568}}
\put (130,235){\makebox(0,0){0.842}}
\put (178,265){\makebox(0,0){0.237}}
\put (178,235){\makebox(0,0){0.706}}
\put (178,220){\makebox(0,0){0.929}}
\end{picture}
\vspace{-6.5cm}
\caption{{\it The values of depth of the effective potential well and the solutions for} ${\rho}$ {\it for} $S$-{\it states} {\it for} $\bar{\varphi}_0 = 8$, ${\xi}_0 = 2.5$ {\it and various values of}  ${\mu}$:
$(1)$  ${\mu} \sqrt{2} = 4.5$,
$(2)$  ${\mu} \sqrt{2} = 5$,
$(3)$  ${\mu} \sqrt{2} = 5.5$.}
\end{figure}
\begin{figure}[hbtp]
\vspace{-1cm}
\begin{picture}(200,400)
\put (10,340){\line(0,-1){40}}
\put (45,340){\line(0,-1){40}}
\put (10,300){\line(1,0){35}}

\put (72.5,340){\line(0,-1){80}}
\put (122.5,340){\line(0,-1){80}}
\put (72.5,260){\line(1,0){50}}

\put (150,340){\line(0,-1){120}}
\put (211,340){\line(0,-1){120}}
\put (150,220){\line(1,0){61}}

\put (10,306){\line(1,0){35}}
\put (10,323){\line(1,0){35}}
\put (72.5,266){\line(1,0){50}}
\put (72.5,286){\line(1,0){50}}
\put (72.5,317){\line(1,0){50}}
\put (150,227){\line(1,0){61}}
\put (150,248){\line(1,0){61}}
\put (150,280){\line(1,0){61}}
\put (150,324){\line(1,0){61}}

\put (25,190){\makebox(0,0){(1)}}
\put (95,190){\makebox(0,0){(2)}}
\put (180,190){\makebox(0,0){(3)}}
\put (0,306){\makebox(0,0){{\scriptsize -3.40}}}
\put (0,323){\makebox(0,0){{\scriptsize -1.66}}}
\put (62.5,266){\makebox(0,0){{\scriptsize -7.35}}}
\put (62.5,286){\makebox(0,0){{\scriptsize -5.42}}}
\put (62.5,317){\makebox(0,0){{\scriptsize -2.53}}}
\put (140,227){\makebox(0,0){{\scriptsize -11.3}}}
\put (140,248){\makebox(0,0){{\scriptsize -9.25}}}
\put (140,280){\makebox(0,0){{\scriptsize -6.02}}}
\put (140,324){\makebox(0,0){{\scriptsize -1.64}}}

\end{picture}
\vspace{-5.5cm}
\caption{\it A sketch of the effective potential well with its respective S-states energy levels for the sets of parameters from Figure 1.}
\end{figure}
The values of ${\rho}$ satisfying Eq.(\ref{simcondition}) can be found graphically. As in the case of the Schr\"odinger equation for a spherical well, the number of bound states depends on the depth and width of the original well potential, i.e. on 
$\bar{\varphi}_0$ and ${\xi}_0$. In addition, the bound state spectrum depends on the rest energy represented by the parameter ${\mu}$. As we can see from Figures 1 and 2, the larger ${\mu}$, the deeper the effective potential well and the more energy levels in it. For given values of $\bar{\varphi}_0$ and ${\xi}_0$, there is always a value of ${\mu}$ given by Eq.(\ref{orig}) for which the bound state spectrum of the Schr\"odinger equation with the potential given by Eq.(\ref{sphwello}) is reproduced. In Figure 2, this is represented by the set of parameters $(2)$.

\begin{figure}
    \centering
    \includegraphics[width=0.5\linewidth]{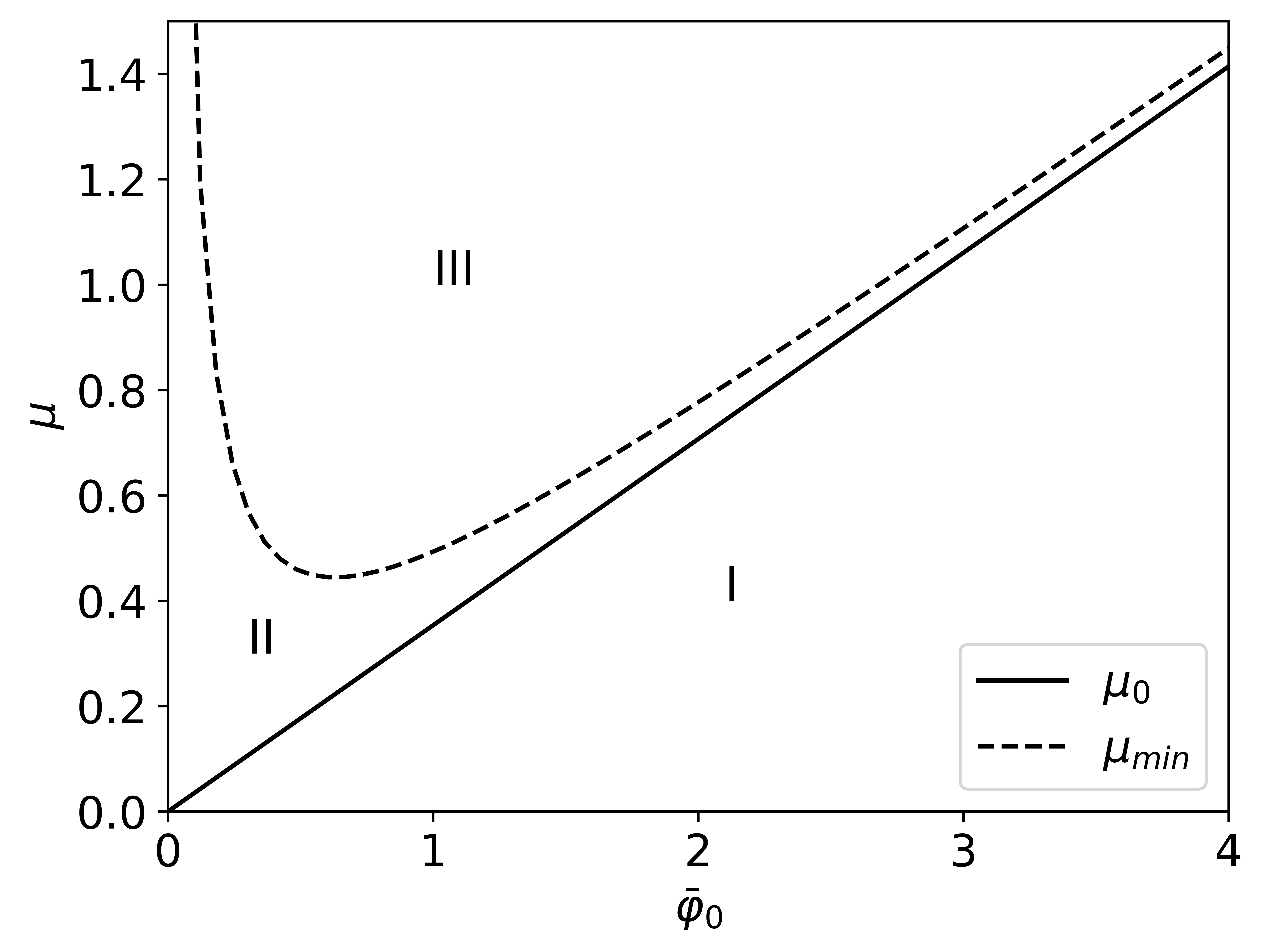}
    \caption{${\mu}-\bar{\varphi}_0$ - phase diagram for the effective potential $V_{\rm eff}$ with ${\xi}_0=2.5$. There are no bound states in phase I, where $V_{\rm eff}$ is a potential step, and in phase II, where $V_{\rm eff}$ is a spherical well potential but the Fermi particle does not have enough rest energy to be bounded. The phase with bound ${\rm S}$-states is phase III.}
    \label{fig:enter-label}
\end{figure}

It can been seen graphically that only when
\begin{equation}
\bar{\xi}_0 \ge \frac{{\pi}}{2}
\end{equation}
or
\begin{equation}
{\mu} \ge {\mu}_{\rm min} \equiv {\mu}_0  + \frac{{\pi}^2}{8 \sqrt{2}  \bar{\varphi}_0 {\xi}_0^2}
\end{equation}
can Eq.(\ref{simcondition}) have a solution. Herein,
${\mu}_{\rm min}$ is the minimum non-zero value of ${\mu}$ to achieve the bound S-states (see Figure 3), and
\begin{equation}
E_{{\rm rest},{\rm min}} \equiv \frac{1}{8} {\left[ \bar{\varphi}_0 + \frac{{\pi}^2}{4\bar{\varphi}_0 {\xi}_0^2} \right]}^2 mc^2
\end{equation}
is the corresponding minimum value of rest energy. For an electron, the estimates of $E_{{\rm rest},{\rm min}}$ are $4.14$ MeV (for $\bar{\varphi}_0 = 8$, ${\xi}_0 = 2.5$) and $0.31$ MeV (for $\bar{\varphi}_0 = 2$, ${\xi}_0 = 2.5$).

\vspace{5 mm}

\section{Discussion}

The LL equation with an external electromagnetic field was known to provide a direct derivation of the Pauli equation. Whereas in the phenomenological approach the spin degree of freedom is introduced "ad hoc", the direct derivation requires only the linearization of the non-relativistic Schr\"odinger equation followed by minimal coupling. The GLL equation does not change this picture. The rest energy term can be removed by redefining the Fermi field components reducing the GLL equation to the LL one.

The non-relativistic counterpart of the Dirac equation should not be considered necessarily as its non-relativistic limit. The GLL equation gives us a new insight into the physical consequences of rest energy. There are external potentials in which the rest energy contribution becomes non-trivial. We can still change the reference point and shift the energy levels by a constant value. However, the rest energy effects are not erased by the shift. With the rest energy included, the GLL equation leads to a physical picture some elements of which cannot be seen in the framework of the LL equation. We have demonstrated this for a non-relativistic Fermi field in a spherical well type potential of finite depth.

In this potential, the spin effects do not show themselves, and the results are the same for spin up and spin down components. This allows us to see clearly the rest energy effects. One of the new elements is the fact that the existence of bound states and their number depend not only on the depth and width of the well, as it was known before, but also on the rest energy of the Fermi particle. The particle entering the region of the well is confined to it if the rest energy differs from zero and exceeds a minimum value. The minimum is determined by the depth and width of the well. Otherwise, the particle is not captured by the well. If it is in a bound state and its rest energy goes below the minimum value, this destroys the bound state, and the particle leaves the well. 

The finite well potentials have multiple applications in condensed matter systems. The spherical well type potentials are used in semiconductor quantum dots. The experimental realization of potentials with the rest energy contribution would provide its impact on bandgap structure and functions of quantum dots. If observed experimentally, the threshold value ${\mu}_{\rm min}$ could be used to estimate the value of the rest mass $m_0$.

The approach presented can be applied to a more general choice of ${\varphi}$ in Eq.(\ref{simplepot}). Then the solution of the GLL equation will exhibit both the rest energy and spin effects including spin-orbit coupling.

\end{document}